\documentclass[aps, pra,reprint,superscriptaddress]{revtex4-1}
\usepackage{graphicx}
\usepackage{amsmath}
\usepackage{soul,xcolor}
\usepackage{color}
\usepackage{placeins}
\usepackage{braket}
\usepackage[normalem]{ulem}
\usepackage{verbatim}
\usepackage{natbib}
\usepackage[normalem]{ulem}

\usepackage{mathtools}

\definecolor{dgreen}{rgb}{0.0, 0.5, 0.0}

\newcommand{\eesr}{$^{88}$Sr~}
\newcommand{\efsr}{$^{84}$Sr~}

\begin{document}

\preprint{APS/123-QED}

\title{Heteronuclear Rydberg molecules}

\author{J.\,D. Whalen}
\author{S.\,K.\,Kanungo} 
\author{Y. Lu}
\affiliation{
 Department of Physics \& Astronomy and Rice Center for Quantum Materials, Rice University, Houston, TX 77251, USA
}

\author{S. Yoshida} 
\author{J. Burgd\"orfer}
\affiliation{%
Institute for Theoretical Physics, Vienna University of Technology, Vienna, Austria, EU
}%

\author{F.\,B. Dunning}
\affiliation{
 Department of Physics \& Astronomy and Rice Center for Quantum Materials, Rice University, Houston, TX 77251, USA
}

\author{T.\,C. Killian}%
\email{corresponding author: killian@rice.edu}
\affiliation{
 Department of Physics \& Astronomy and Rice Center for Quantum Materials, Rice University, Houston, TX 77251, USA
}
\date{\today}

\begin{abstract}
We report the creation of heteronuclear ultralong-range Rydberg-molecule dimers by excitation of minority \eesr atoms to $5sns\,^3S_1$ Rydberg states ($n=$ 31--39) in a dense background of $^{84}\text{Sr}$. We observe an isotope shift of the $\nu=0$ vibrational state over this range of $n$ and compare our measurements with a theoretical prediction and a simple scaling argument. At low principal quantum number the isotope shift is sufficiently large to produce heteronuclear dimers with almost perfect fidelity. When the spectral selectivity is limited, we obtain a lower bound on the ratio of heteronuclear to homonuclear excitation probability of 30 to 1 by measuring the scaling of the molecular excitation rate with varying relative densities of \eesr and \efsr in the ultracold mixture.
\end{abstract}

\maketitle

Ultralong-range Rydberg molecules (RMs) are formed by the scattering between an excited Rydberg electron and at least one nearby neutral atom in a dense gas.
Initially predicted theoretically \cite{PhysRevLett.85.2458}, these molecules have been the subject of intense study following their initial observation in Rb \cite{bendkowsky2009observation} and subsequent observations in Cs \cite{Booth99} and Sr \cite{PhysRevA.92.031403}.
The internuclear potential inherits its shape from the Rydberg-electron probability distribution and supports bound states with bond lengths on the order of the radius of the Rydberg orbital $R_n \sim 2n^2$, $\sim 100\text{ nm}$ at $n=35$.
Recent studies have focused on the large permanent electric dipole moments of so-called trilobite and butterfly RMs \cite{Booth99, niederprum2016observation}, ultracold chemistry and the stability of RMs in dense cold gases \cite{PhysRevA.96.042702,PhysRevX.6.031020}, and on methods to use RMs to study low energy atom-ion scattering \cite{PhysRevLett.120.153401} and probe the pair-correlation function of quantum gases \cite{PhysRevA.100.011402}.

In this work we demonstrate photo-excitation of heteronuclear RM dimers using a mixture of \eesr and ${}^{84}\text{Sr}$.
We observe an isotope shift in the binding energy between the ground vibrational state of $^{84}\text{Sr}+{}^{84}\text{Sr}$, $^{88}\text{Sr}+{}^{84}\text{Sr}$ and $^{88}\text{Sr}+{}^{88}\text{Sr}$ RMs, and measure the scaling of molecule production with \eesr and \efsr density.
Spectroscopy of heteronuclear molecules provides a sensitive probe of RM potentials and can be used to measure the relative densities of constituent atomic species in a mixture, as proposed in theoretical work on bialkali heteronuclear RMs \cite{PhysRevA.98.042706}.
Excitation of heteronuclear RM dimers from varied and well-controlled atomic constituents will enable the study of spatial correlations and collisional wavefunctions in atomic mixtures \cite{PhysRevA.100.011402}.
The combination of \eesr and \efsr is particularly interesting in this context because the system possesses an extremely large scattering length, $a_{88-84} \sim 1800 \, a_0$, where $a_0$ is the Bohr radius \cite{PhysRevA.78.042508,PhysRevA.78.062708}.

\begin{figure}
\begin{center}
\includegraphics[width=\columnwidth]{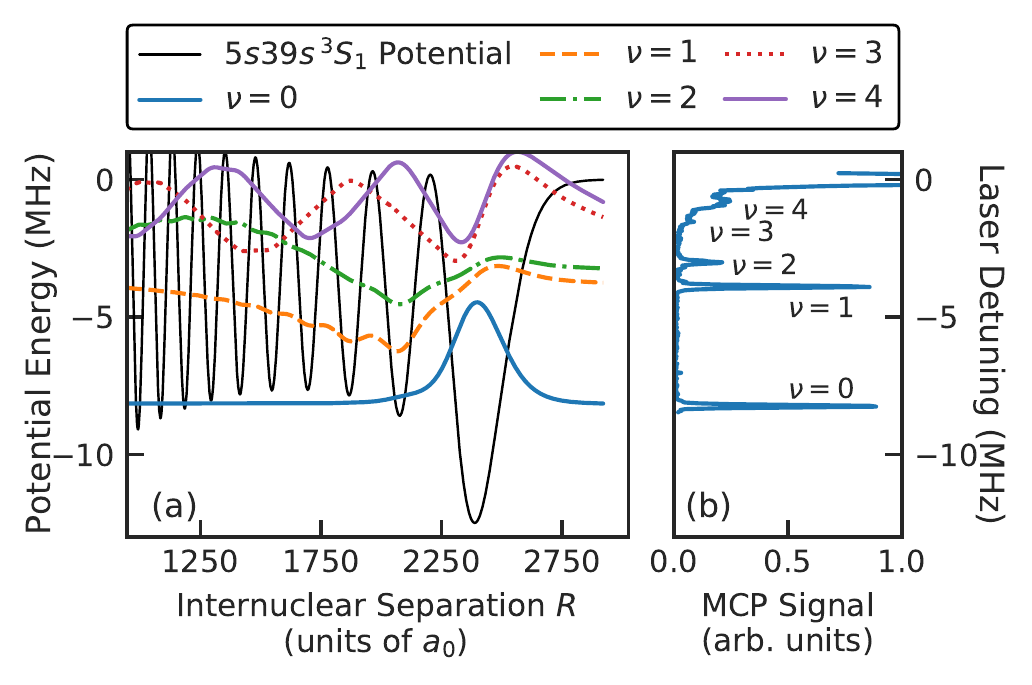}
\caption{a) The potential formed by the scattering between the $5s39s\,^3S_1$ Rydberg electron and a neutral Sr atom. The bound state wavefunctions are labeled by their vibrational quantum numbers $\nu$ and are offset by their binding energy. b) RM excitation spectrum observed when the excitation laser is detuned from atomic resonance (see text).\label{fig:rydberg_potential}}
\end{center}
\end{figure}

The interaction of an excited Rydberg electron and a ground-state atom gives rise to a potential similar to that shown in Figure \ref{fig:rydberg_potential}.
 The potential experienced by a ground-state atom, at distance $R$ from the Rydberg atom nucleus, can be described in the Born-Oppenheimer approximation by a modified \hbox{Fermi pseudopotential \cite{PhysRevLett.85.2458,fermi1934fermi,omont},}
\begin{equation}\label{eq:potential}
V(R) = \frac{2\pi\hbar^2 A_s(k)}{m_e}\left|\psi_{ns}(R)\right|^2 + \frac{6\pi\hbar^2 A_p^3}{m_e}\left|\nabla\psi_{ns}(R)\right|^2,
\end{equation}
where $\psi_{ns}(R)$ is the Rydberg electron wavefunction, $m_e$ is the electron mass, and $A_s(k)$ and $A_p$ are s-wave and p-wave scattering lengths, respectively. For simplicity momentum ($k$) dependence of the scattering length is
only included for the s-wave interaction.
For strontium $A_s(0)=-13.3\,a_0$ and $A_p = 9.7\,a_0$ \cite{PhysRevA.92.031403} giving rise to an attractive potential. (In alkali atoms such as Rb the p-wave interaction \cite{PhysRevLett.116.053001} and the hyperfine structure \cite{PhysRevLett.117.123002,PhysRevLett.118.223001} affect the RM potential. In the strontium isotopes used here, these effects are absent.)
 This attractive potential supports a manifold of vibrational states labeled by the quantum number $\nu$ that appear as resonances red detuned from the atomic Rydberg state as shown in Figure \ref{fig:rydberg_potential}.
The binding energies of these molecular states scale as $n^{-6}$.
For each principal quantum number, the potential supports many bound states, but we focus on the highly localized $\nu=0$ state for $n=31-39$ here. The polarization potential between the background atom and ion core is negligible for these states.

In the present experiment we create $\nu=0$ homonuclear RM dimers comprised of ${}^{88}\text{Sr}^{*}+{}^{88}\text{Sr}$ and ${}^{84}\text{Sr}^*+{}^{84}\text{Sr}$, and heteronuclear dimers of ${}^{88}\text{Sr}^*+{}^{84}\text{Sr}$, where the star indicates the atom in the Rydberg state.
Homonuclear dimers are excited from ultracold samples of either pure \eesr or pure ${}^{84}\text{Sr}$.
The chosen isotope is laser cooled in a broadband magneto-optical trap (MOT) operating on the $5s^2\,^1S_0\rightarrow5s5p\,^1P_1$ cycling transition at 461 nm, from which a fraction of the atoms spontaneously decay into the metastable $5s5p\,^3P_2$ state and are magnetically trapped in the quadrupole field of the MOT \cite{PhysRevA.67.011401}. The atoms are repumped to the ground state and further laser cooled using a narrow-band MOT operating on the $5s^2\,^1S_0\rightarrow 5s5p\,^3P_1$ intercombination line at 689 nm and are then loaded into an optical dipole trap (ODT) formed by two perpendicular light sheets at 1064 nm as described elsewhere \cite{molphys}.
We reduce the intensity of the ODT beams over an interval of 500 ms to allow for evaporative cooling after which \efsr reaches a temperature of $\sim800$ nK and a density of $9 \times 10^{12}\,\text{cm}^{-3}$. The weak interactions between \eesr atoms (s-wave scattering length $a=-2\,a_0$ \cite{PhysRevA.78.062708,stein20101}) make evaporative cooling less efficient, resulting in a final temperature of 1.1 $\mu$K and density of $6 \times 10^{12}\,\text{cm}^{-3}$ for \eesr at the same trap depth.

To create heteronuclear RMs we prepare an ultracold mixture of \eesr and ${}^{84}\text{Sr}$. The two isotopes are sequentially laser cooled in the broadband MOT and loaded into the magnetic trap.
From the magnetic trap the atoms are simultaneously repumped and cooled in a dual-isotope narrow-band MOT using laser light tuned to each isotope.
The isotope shift of the $5s5p\,^3P_1$ intercombination line is sufficiently large that the \eesr and \efsr MOTs do not interfere with each other.
From the narrow-line MOT, the mixture is loaded into the ODT and evaporatively cooled to the same trap depth as above.
The sample temperature is set to $\sim 750\text{ nK}$, and the large interspecies scattering length and resulting high collision rate ensures that the mixture is in thermal equilibrium.
We can selectively measure the number and temperature of each isotope using absorption imaging on the $5s5p\,^1P_1$ line and find that the temperatures of the two isotopes are equal to within 10\% at the end of the evaporation.
During the evaporation stage there is rapid atom loss due to the large three-body recombination rate associated with the strong interspecies interactions, which limits the maximum attainable density.
We work with a relatively low peak density of $5\times10^{12}\text{ cm}^{-3}$, which yields samples with lifetimes on the order of a few hundred milliseconds.
We can adjust the relative densities of the two isotopes by varying the load times of the broadband MOT. The ratio of \eesr to \efsr density is varied from 0.03 to 1 while maintaining approximately the same total number density and atom number $N\sim 3\times10^5$.

RMs are created in these ultracold samples using a two-photon excitation comprised of a fixed-wavelength 689 nm photon tuned near the $5s5p\,^3P_1$ level of the selected isotope and a 320 nm photon tuned to select the target Rydberg state. The target atom is excited to the $5sns\,^3S_1$ state with $n=31-39$. For excitation of homonuclear \eesr(${}^{84}\text{Sr}$) molecules, we apply a 689 nm photon detuned 15 MHz (80 MHz) from the $5s5p\,^3P_1$ state. To excite heteronuclear molecules we selectively excite \eesr to the $5sns\,^3S_1$ state in a background of \efsr using the same laser detunings used to create homonuclear \eesr molecules. The isotope shift of the $5sns\,^3S_1$ level between \eesr and \efsr is 445 MHz over the range of principal quantum numbers considered in this work and ensures complete isotope selectivity of the atom that is excited electronically. The detuning of the 689 nm photon from the $5s5p\,^3P_1$ level of \efsr is sufficiently large that off-resonant scattering does not cause heating of the sample.

\begin{figure}
\begin{center}
\includegraphics[width=\columnwidth]{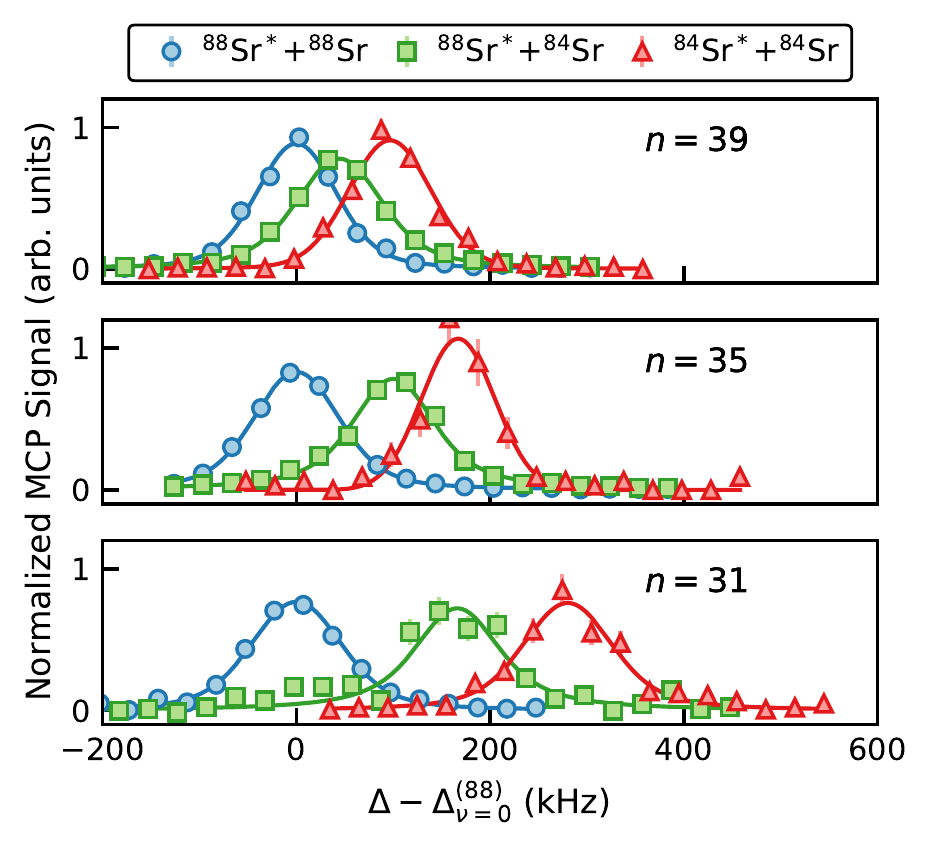}
\caption{Characteristic excitation spectra of $\nu=0$ RMs in samples of pure \eesr (blue circles), minority \eesr excited in majority \efsr (green squares), and pure \efsr (red triangles) for principal quantum numbers $n=31-39$. Each spectrum is plotted against the detuning, $\Delta$, from the atomic Rydberg line of the excited isotope, shifted by the detuning for the ${}^{88}\text{Sr}^*+{}^{88}\text{Sr}$ $\nu=0$ dimer state, $\Delta_{\nu=0}^{(88)}$. Resonances for molecules with smaller reduced mass are shifted closer to the atomic line as expected. All spectra are taken in the same trap and with similar total densities (see text). For the heteronuclear spectra the ratio of the \eesr and \efsr densities is $\sim0.1$. \label{fig:spectrum}}
\end{center}
\end{figure}

We generate spectra such as those shown in Figures \ref{fig:rydberg_potential} and \ref{fig:spectrum} by scanning the energy of the 320 nm photon. We apply a 20 $\mu$s excitation pulse followed by an electric field ramp that ionizes any excited Rydberg atoms or molecules present in the sample. The liberated electrons are guided to a microchannel plate (MCP) detector where they are detected and counted. We repeat the excitation process 500 times per sample over 100 ms to build up statistics. The ODT beams are turned off during excitation to avoid AC stark shifts. To preclude the effects of Rydberg-Rydberg interactions we lower the power of the excitation lasers to ensure that fewer than one Rydberg atom is created per excitation pulse on average.

The first evidence for production of heteronuclear Rydberg molecules is provided by the presence of an isotope shift in the binding energy of the $\nu=0$ dimer state.
In Figure \ref{fig:spectrum} we show normalized excitation spectra for $\nu=0$ RM dimers in three different samples: a pure \eesr sample, excitation of a minority \eesr in a  majority \efsr background, and a pure \efsr sample for principal quantum numbers between $n=31-39$. Spectra are plotted versus the detuning from the atomic Rydberg line of the electronically excited isotope, shifted by the detuning corresponding to the $\nu=0$ line for ${}^{88}\text{Sr}^*+{}^{88}\text{Sr}$. Resonances for molecules with lighter reduced mass are shifted closer to the atomic line as is expected.

The isotope shift increases with decreasing principal quantum number with a scaling that can be understood with a simple model that is valid for the range of quantum numbers studied here.
The molecular potential is approximated as harmonic
around $R = R_c$ at which the RM potential has a minimum value $V(R_c) = -d$, i.e.,
\begin{equation}
V(R) \simeq \frac{1}{2} \mu \omega^2 (R - R_c)^2 - d
\end{equation}
with $\mu$ the reduced mass of the dimer pair.
Because the $\nu=0$ state is highly localized in the most outer well (see Fig. \ref{fig:rydberg_potential}),
the binding energy can be approximated by $E_{\nu=0} = d - \hbar\omega/2$.
Mass-dependent effects on the RM potential, such as variation of the reduced mass of the electron and background atom are negligible on the level of our measurement accuracy. Effects beyond the Born-Oppenheimer approximation,
such as the mass polarization energy between the Rydberg electron
and the ground-state atom, are also negligible compared to the isotope
shifts.
Therefore the shift in the binding energy is related to the shift in the characteristic frequency $\omega$, which varies as $\mu^{-1/2}$, where $\mu$ is the reduced mass of the dimer pair.
To first order the relative shift in the harmonic potential's frequency is
proportional to $\omega$
\begin{equation}
\frac{\Delta\omega}{\omega} = -\frac{1}{2}\frac{\Delta\mu}{\mu} \, .
\end{equation}

The expected scaling of $\omega$ and the isotope shift with $n$ can be estimated within the harmonic approximation.
The potential felt by the neutral atom follows the Rydberg electron probability density distribution for the $5sns\,^3S_1$ state and therefore
has $N \simeq n$ nodes.
The width, $w$, of the outer well scales as $w \sim R_n/N \sim 2n^2/n \sim n$, where $R_n$ is the radius of the Rydberg orbital.
The depth of the outer well scales as $d \sim n^{-6}$, the inverse of the volume of the electron orbital \cite{bendkowsky2009observation}.
The potential parameters can thus be approximated through $\mu \omega^2 w^2 \sim d$, implying $\omega \sim \sqrt{d/w^2} \sim n^{-4}$.
The isotope shift is proportional to $\omega$ as discussed above, and therefore should also scale as $n^{-4}$.
Inclusion of anharmonic corrections \cite{morsefeshbach}
results in a somewhat more rapid decay
with $n$, $n^{-\alpha}$, with $\alpha > 4$. Numerical calculations confirms
such an approximate scaling.
For the range of principal quantum numbers considered in this work the harmonic approximation yields an estimated isotope shift from a few hundred to a few tens of kHz which is close to the measured values.
For quantum numbers $n\gtrsim40$, the $\nu=0$ vibrational state extends to adjacent wells of the potential to an increasing degree, and the harmonic approximation becomes poor.

\begin{figure}
\begin{center}
\includegraphics[width=\columnwidth]{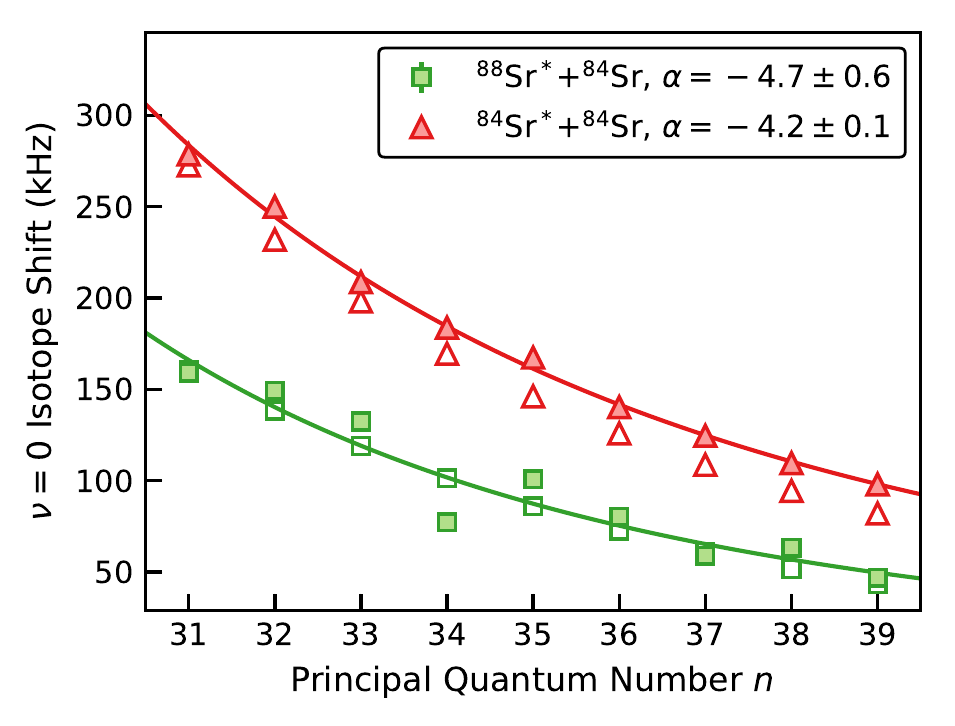}
\caption{The measured (filled symbols) and calculated (open symbols) isotope shifts for the $\nu=0$ RM dimer in $^{88}\text{Sr}^*+{}^{84}\text{Sr}$ and $^{84}\text{Sr}^*+{}^{84}\text{Sr}$ compared to $^{88}\text{Sr}^*+{}^{88}\text{Sr}$. The measured data are fitted to a power law $(n-\delta)^{\alpha}$ where $\delta=3.371$ is the quantum defect of the $5sns\,^3S_1$ state. The fitted $\alpha$ are shown in the legend and agree reasonably with the expected $n^{-4}$ scaling (see text).\label{fig:shift_scaling}}
\end{center}
\end{figure}

Figure \ref{fig:shift_scaling} shows the isotope shifts of spectral-line centers for
${}^{88}\text{Sr}^*+{}^{84}\text{Sr}$ and ${}^{84}\text{Sr}^*+{}^{84}\text{Sr}$ RMs with respect to ${}^{88}\text{Sr}^*+{}^{88}\text{Sr}$. Statistical uncertainties for experimental data are smaller than the symbol size. The results agree reasonably well with
the expected $n^{-4}$ scaling. 
Figure \ref{fig:shift_scaling} also shows theoretical predictions, 
which are calculated using  the formalism of \cite{Ding_2019} to model RM excitation spectra. For this calculation, molecular energy levels are obtained by numerical diagonalization
of the Hamiltonian matrix formed using the molecular potential (Eq. \ref{eq:potential}) with electron-atom scattering lengths from \cite{PhysRevA.92.031403}.
A thermal average over  initial collision energies and sum over  partial waves up to $\Lambda=3$ is performed using $\Lambda$-dependent Frank-Condon factors.
For $n = 31$, the RM rotational constant is $\sim 20$\,kHz and the spectral line can shift several tens of kHz  from the position predicted in the absence of rotational effects. This shift is isotope dependent because heteronuclear RM dimers are excited from a pair of distinguishable bosons and states with both even and odd rotational quantum number can be excited, while only even values of $\Lambda$ contribute in the homonuclear case.  In addition, the very large scattering length $a_{84-88}$  suppresses the Franck-Condon factors for s-waves for ${}^{88}\text{Sr}^*+{}^{84}\text{Sr}$.
 The quantitative agreement between isotope-shift values for experiment and theory is reasonable given the lower resolution of measurements used to determine parameters of the molecular potential \cite{PhysRevA.92.031403}.

The presence of the isotope shift discussed above is conclusive evidence of the creation of heteronuclear RMs. For future applications one key quantity of interest is the fidelity with which heteronuclear RMs can be produced. It is clear from Figure \ref{fig:spectrum} that for lower quantum numbers ($n\lesssim35$) and with a sufficiently narrow excitation-laser linewidth, one can excite heteronuclear molecules with almost perfect fidelity by taking advantage of the spectroscopic resolution of the isotope shift. However, when the involved electronic state, the combination of constituent isotopes, or the excitation laser system does not permit the required spectral selectivity, another method of determining excitation fidelity is required.

As an example we consider the excitation of heteronuclear RMs at $n=39$ where the isotope shift is 48 kHz, which is about half of the linewidth of the Rydberg excitation limited by the linewidth of the 320 nm laser system. If we assume that the distribution of atom positions is governed by Poissonian statistics and that the total number density, $n$, is much less than one per Rydberg-orbital volume, the RM excitation rate from the excitation of \eesr in a ${}^{88}\text{Sr}+{}^{84}\text{Sr}$ mixture as a function of laser frequency $\nu$ is
\begin{multline}\label{eq:nur}
R(\nu) = \int d^3r\, \kappa\left(
	n_{88}(r)^2 \mathcal{F}(\nu-\nu^{(88-88)}_0) + \cdots \right.\\
	\left. n_{88}(r) n_{84}(r)\mathcal{F}(\nu-\nu_0^{(88-84)})\right)
\end{multline}
where $n_x(r)$ is the number density distribution for each isotope, $\mathcal{F}(\nu-\nu_0)$ is a lineshape function with unit integral centered at laser frequency $\nu_0$, and $\kappa$ is a rate constant that is a function of isotope-independent quantities such as electronic matrix elements, laser power, and other experimental parameters that are held fixed here. The Bose-enhancement of the homonuclear RM excitation rate \cite{PhysRevA.100.011402} is about 10\% for $n=39$ at these temperatures and is neglected in our analysis.

Frequency integration over the line shape gives
the integrated rate, $\mathcal{S}$,
\begin{equation}
\mathcal{S}=\int d\nu\, R(\nu) =
  \kappa( n_{0,88}^2 + n_{0,88}n_{0,84}) \, V.
\end{equation}
For a harmonic trap the effective volume that arises from the integration of the spatial density distribution is
\begin{equation}
V = \left(\frac{\pi k_B T}{m \bar{\omega}^2}\right)^{3/2}
\end{equation}
with geometric mean of the trap frequencies $\bar{\omega}$
and
the spatial density distribution $n_x(r)$ is assumed
to be a Gaussian with the peak density $n_{0,x}$ for each isotope. The effective volume is, in principle, isotope dependent, but when the effective volumes for each isotope are calculated we find that they are equal to within 10\% over all of the densities considered and we assume these volumes as equal.
We normalize the integrated rate $\mathcal{S}$ by $n_{0,88}^2 V$ to obtain a relation that is a function of the ratio of the two isotope densities.
\begin{equation}\label{eq:density_scaling}
\frac{\mathcal{S}}{V\,n_{0,88}^2} =
\kappa \left( 1 + \frac{n_{0,84}}{n_{0,88}} \right)
\, .
\end{equation}

\begin{figure}
\begin{center}
\includegraphics[width=\columnwidth]{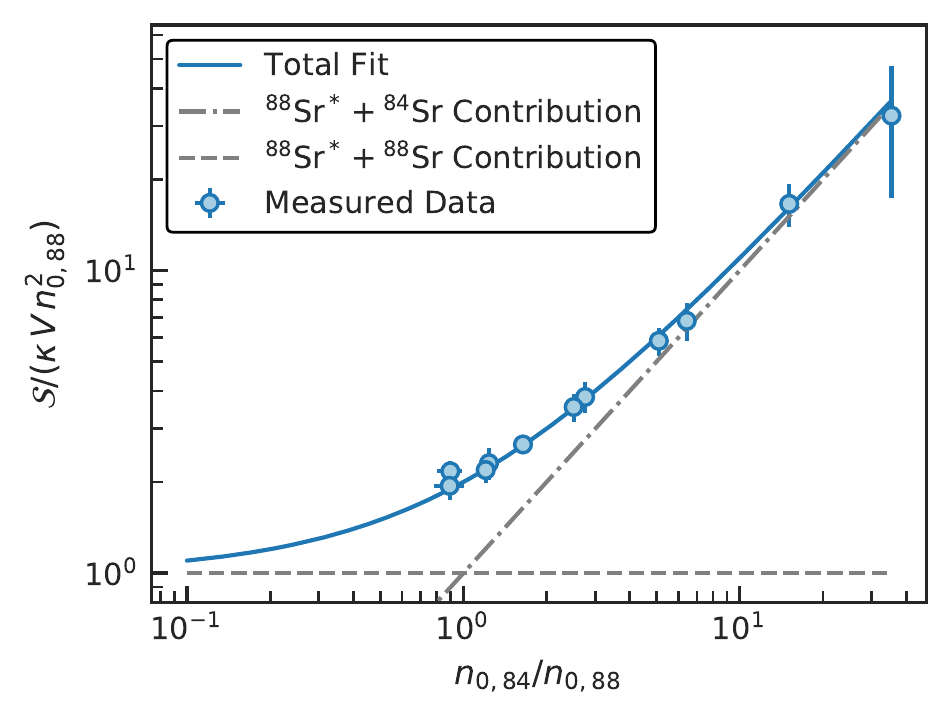}
\caption{The scaling of normalized Rydberg molecule signal with density. The blue points are integrated Rydberg molecule signals normalized by the density of \eesr and and fitted to the scaling relation given in Equation \ref{eq:density_scaling} with a single constant of proportionality (blue line). The grey dashed and dot-dashed lines represent the individual contributions to the signal from ${}^{88}\text{Sr}+{}^{88}\text{Sr}$ pairs and ${}^{88}\text{Sr}+{}^{84}\text{Sr}$ pairs respectively. It is clear that both terms are required to accurately explain the behavior of the molecular excitation rate. \label{fig:density}}
\end{center}
\end{figure}

We obtain RM excitation spectra in ${}^{88}\text{Sr}+{}^{84}\text{Sr}$ mixtures where we vary the ratio of \eesr density to \efsr density from 0.03 to 1 as discussed above. We scan the laser detuning over a range that includes both the ${}^{88}\text{Sr}^* + {}^{88}\text{Sr}$ and ${}^{88}\text{Sr}^*+{}^{84}\text{Sr}$ lines and integrate the spectrum. The integrated spectra are normalized according to Equation \ref{eq:density_scaling} and fitted with a single constant, $\kappa$. As shown in Figure \ref{fig:density}, the measured data are well described by this scaling relation. For the lowest \eesr fraction, we have a heteronuclear production fidelity of about 30 to 1. This fidelity is only a lower bound, however, as we have integrated over the entire spectrum and neglected any spectral selectivity. We can easily improve the heteronuclear excitation fidelity if spectral resolution is utilized, even for $n=39$.

In summary, we have observed the excitation of heteronuclear Rydberg molecules in a mixture of ${}^{88}\text{Sr}+{}^{84}\text{Sr}$. The presence of an isotope shift provides clear evidence of the creation of heteronuclear molecules, and the dependence of the integrated RM spectrum on the relative densities is well described by a model incorporating heteronuclear and homonuclear production. When the isotope shift is large compared to the spectral linewidth, essentially pure samples of heteronuclear molecules can be created. In the limit of unresolved lines, we have demonstrated homonuclear impurity below 1 part in 30. Heteronuclear RMs present many new opportunities for studying quantum systems such as probing interparticle and interspecies pair correlation functions in multi-component many body systems \cite{venegasgomez2020adiabatic,PhysRevA.101.013603} using the methods of \cite{PhysRevA.100.011402}. We can probe the scattering wavefunction of a strongly interacting mixture such as the ${}^{88}\text{Sr}+{}^{84}\text{Sr}$ combination used in this work. Such an \textit{in situ} measurement of the pair-correlation function in the presence of strong interactions can be related to the Tan contact \cite{tan2008energetics,PhysRevA.86.053633}, and the temporal resolution available with RM excitation could provide a probe of the dynamics of the contact after an interaction quench. This new variety of RMs opens paths to explore the sensitivity of the isotope shift to the mass polarization energy for Rydberg trimers or larger molecules, and RMs involving high $L$ Rydberg states. In lighter atomic species, e.g. Li, the isotope shift is expected to be much larger than observed in this work \cite{PhysRevLett.120.153401}. Heteronuclear RMs could also enable studies of new types of Rydberg impurities in quantum gases \cite{PhysRevResearch.2.023021}. Very recent work in bialkali mixtures of K and Cs have shown that heteronuclear RMs can be used to benchmark theoretical models of electron-atom scattering  \cite{peper2020heteronuclear}.

\textbf{Acknowledgments:} This research was supported by the NSF under Grant No. 1904294, the AFOSR under Grant No. FA9550-12-1-0267, the Robert A. Welch Foundation under Grants Nos. C-0734 and C-1844, the FWF (Austria) under Grants Nos. FWF-SFB041ViCom and FWF-doctoral
college W1243. The Vienna Scientific Cluster was used for the calculations.

\bibliography{hetero_molecules}

\end{document}